\documentclass[aps,reprint]{revtex4-1}
\usepackage[dvipdfmx]{graphicx,color}
\usepackage{dcolumn}
\usepackage{bm}
\usepackage{amsmath,amsthm,amssymb}
\usepackage{mathrsfs}
\usepackage{gt15}

\begin{document}

\title{
Effective Hamiltonian theory for nonreciprocal light propagation in magnetic Rashba conductor
}

\author{Hideo Kawaguchi$^{1,2}$}
\email[Email address: ]{hideo.kawaguchi@riken.jp}
\author{Gen Tatara$^2$}
\affiliation{$^1$ Graduate School of Science and Engineering, Tokyo Metropolitan University, Hachioji, Tokyo 192-0397, Japan \\
$^2$ RIKEN Center for Emergent Matter Science (CEMS), 2-1 Hirosawa, Wako, Saitama 351-0198, Japan}

\date{\today}

\begin{abstract}
Rashba spin-orbit interaction leads to a number of electromagnetic cross-correlation effects by inducing a mixing of electric and magnetic degrees of freedom. 
In this study, we investigate the optical properties of a magnetic Rashba conductor by deriving an effective Hamiltonian based on an imaginary-time path-integral formalism. 
We show that the effective Hamiltonian can be described in terms of toroidal and quadrupole moments, as  has been argued in the case of insulator multiferroics. 
The toroidal moment turns out to coincide with the spin gauge field induced by the Rashba field.
It causes Doppler shift by inducing intrinsic spin current, resulting in anisotropic light propagation (directional dichroism) irrespective of the polarization. 
The quadrupole moment on the other hand  results in a magneto-optical phenomenon such as a Faraday effect for circularly polarized waves. 
\end{abstract}


\maketitle

\section{Introduction: Rashba-induced electromagnetic cross-correlation effects}  \label{Intro}

Conversion between electric signals and magnetic information plays an important role in the development of currently available information technologies.
Such conversions were first performed using classical laws of electromagnetism, such as Amp{\'e}re's law and Faraday's law. 
However, these classical mechanisms have not been able to sufficiently meet the recent technological requirements of fast processing of large amounts of information and high-density storage; hence, they have been gradually replaced by solid-state mechanisms such as spin-transfer torque \cite{Slonczewski96,Berger96}. 
Spin-orbit interaction, which couples the orbital motion of an electron to its spin via a relativistic effect, plays an important role in this context. 
\\
\indent
Recent studies have shown that this interaction becomes prominent for surfaces and interfaces containing heavy metals as it significantly modifies their electric and magnetic properties as a consequence of inversion symmetry breaking \cite{Ast07}. 
The most typical spin-orbit interaction lacking inversion symmetry is the Rashba interaction \cite{Rashba60}, whose Hamiltonian is as shown below.
\begin{align}
\cal{H}_{\rm R} &= i \alphaRv \cdot (\bm{\nabla} \times \bm{\sigma}), 
\end{align}
where $\bm{\sigma}$ is the vector of Pauli matrices and $\alphaRv$ is the Rashba field representing the strength and direction of the Rashba spin-orbit interaction.
This form of interaction is derived directly from the Dirac equation as a relativistic effect, but its magnitude can be significantly enhanced for solids containing heavy elements as compared to that for the vacuum case. 
\\
\indent
A direct consequence of the Rashba interaction is an electromagnetic cross-correlation effect where a magnetization and an electric current are induced by external electric and magnetic fields, $\Ev$ and $\Bv$, as $\Mv=\gamma_{ME}({\bm{\alpha}}_{\rm R}\times\Ev)$ and $\jv=\gamma_{jB}({\bm{\alpha}}_{\rm R}\times\Bv)$, where $\gamma_{ME}$ and $\gamma_{jB}$ are coefficients that generally depend on frequency. 
The emergence of spin accumulation from the applied electric field, mentioned in Ref. \cite{Rashba60}, was studied in detail by Edelstein \cite{Edelstein90}; hence, this effect is sometimes referred to as the Edelstein effect. The generation of electric current by a magnetic field or magnetization, called the inverse Edelstein effect \cite{Shen14}, was recently observed in a multilayer structure consisting of Ag, Bi, and a ferromagnet \cite{Sanchez13}. 
\\
\indent
Furthermore, a recent study showed that the cross-correlation effects of the Rashba spin-orbit interaction lead to anomalous optical properties \cite{Shibata16}. It was shown that the interaction leads to a strongly anisotropic light propagation, resulting in a hyperbolic metamaterial \cite{Narimanov15} that exhibits a negative refraction and a focusing effect. In addition, if the system is magnetic or under the effect of an external magnetic field, directional dichroism, a form of anisotropic wave propagation, has been shown to occur. The directional dichroism was found to be governed by the relative direction of the wave vector and another vector $\bm{{\cal A}}_{\rm R}\equiv \alphaRv\times \Mv$, where $\Mv$ is the magnetization vector. The latter vector is known to be an effective gauge field coupled with the electron's spin (spin gauge field), which generates a spin current \cite{Takeuchi12,Kim12,Nakabayashi14}. From the symmetry point of view, the Rashba field $\alphaRv$ is equivalent to an electric polarization $\Pv$ (Ref.  \cite{Hayami14}); hence, the vector $\bm{{\cal A}}_{\rm R}$ works as a toroidal moment $\tv\equiv \Pv\times\Mv$. The toroidal moment has been reported to act as an effective vector potential for light in the case of multiferroics \cite{Sawada05}; however, microscopic justification for the same has not been provided.
The study in Ref. \cite{Shibata16} further discussed that the effective theory describing magnetic Rashba conductors is similar to the one describing insulator multiferroics. 
Cross-correlation effects were also discussed from the viewpoint of optical chirality in Ref. \cite{Proskurin16}. 
\\
\indent
In this study, we examine the propagation of electromagnetic waves in magnetic Rashba conductors based on an effective Hamiltonian analysis on a microscopic ground. 
We show that the effective Hamiltonian describing the directional dichroism consists of two terms, one representing the Doppler shift and the other denoting the cross-correlation effect induced by a quadrupole moment. The results of our study confirm with those of Ref. \cite{Shibata16} obtained by calculating an optical conductivity. 
\\
\indent
As has been reported by Volovik \cite{Volovik87}, the original spin gauge field (adiabatic spin gauge field) occurs in the absence of Rashba interaction in the strong $sd$-coupling regime \cite{GT14}. 
The coupling of the adiabatic spin gauge field to the electromagnetic field was studied based on the effective Hamiltonian study in Refs. \cite{Kohno09,Kawaguchi14}. It was found that the spin-transfer effect is described using the linear coupling term of the adiabatic spin gauge field and the electric field. 
Reference \cite{Taguchi12} reported the effect of the adiabatic spin gauge field on nonlinear optical effects and a topological inverse Faraday effect was shown to have occurred from a spin Berry's curvature in the absence of a spin-orbit interaction.

\section{Phenomenological argument for electromagnetic property} \label{II}
The existence of cross-correlation effects of Rashba conductors indicates that the electric and magnetic fields, $\Ev$ and $\Bv$, are linearly coupled.
There are two possibilities for this form of interaction: The first one is proportional to $\Ev\cdot\Bv$, and the second one is proportional to their vector product  $\Ev\times\Bv$, as shown below.
\begin{align}
\cal{H}_{\theta}&= \theta\Ev\cdot\Bv, \label{EdotB} \\
{\cal{H}}_{u}&=\uv\cdot(\Ev\times\Bv), \label{ExB}
\end{align}
where $\theta$ is a constant and $\uv$ is a constant vector. 
The first scalar interaction (\ref{EdotB}) does not modify the equation of motion but has a topological effect, and the emergence of such an interaction is restricted to surfaces or interfaces with nontrivial topological properties. 
For instance, such a scalar coupling leads to a mixing of $\Ev$ and $\Bv$ when the topological number $\theta /2\pi$ has a jump at the interfaces \cite{Qi11}. 
(In Sec. \ref{V}, the cross-correlation effect induced by the $\theta$ term in Weyl semimetal is briefly discussed.) 
In contrast, the coupling of the vector product shown in Eq. (\ref{ExB}) may occur in ordinary materials if a vector $\uv$ exists because of the breaking of both spatial-inversion symmetry and time-reversal invariance.
This term, Eq. (\ref{ExB}), has not been discussed in the context of high-energy physics because it is not invariant under the Lorentz transformation.
\\
\indent
It is known that the vector $\frac{1}{\mu}\Ev\times\Bv$ (where $\mu$ is the magnetic permeability of solids) is the Poynting vector representing the momentum of the electromagnetic wave \cite{Jackson98}. The vector interaction of Eq. (\ref{ExB}) can thus be considered to be of the form representing the Doppler shift $\uv\cdot\kv$, where $\kv$ is the wave vector of the electromagnetic wave. 
In fact, in terms of the photon operators $a_{\kv}$ and $a_{\kv}^{\dagger}$, the coupling of Eq. (\ref{ExB}) modifies the photon Hamiltonian as 
\begin{align}
H_{\rm photon}&=\sum_{\kv}(ck-\bm{u}\cdot\kv)a_{\kv}^{\dagger} a_{\kv},
\end{align}
where $c$ is the light velocity. The vector coupling of Eq. (\ref{ExB}) is thus expected to occur when the medium has an intrinsic flow with a velocity proportional to $\uv$.
\\ 
\indent
The Doppler shift picture can also be justified at the level of the equation of motion. The following description shows how electromagnetism is modified by the vector interaction of Eq. (\ref{ExB}).
Two of Maxwell's equations become 
\begin{align}
 \bm{\nabla}\cdot\Ev =& \frac{\rho}{\ez}- \frac{1}{\ez}\bm{\nabla}\cdot(\uv\times\Bv), 
\nonumber \\
 \bm{\nabla}\times\Bv =& \muz\jv+\ez\muz\frac{\partial \Ev}{\partial t} +\muz\frac{\partial }{\partial t}(\uv\times\Bv)-\muz\bm{\nabla}\times(\uv\times\Ev),
\end{align}
whereas the other two remain unchanged ($\bm{\nabla}\cdot\Bv=0$ and $\bm{\nabla}\times \Ev=-\frac{\partial \Bv}{\partial t}$). Here, $\rho$ represents charge density, $\bm{j}$ is the charge-current density, and $\ez$ and $\muz$ are the electric permittivity and magnetic permeability of the vacuum, respectively.
The total electric and magnetic fields can be represented as follows. 
\begin{align}
 \Ev_{\rm tot} =& \Ev+\frac{1}{\ez}(\uv\times\Bv), \nnr
 \Bv_{\rm tot}= & \Bv+\muz(\uv\times\Ev). \label{EBcross}
\end{align}
These relations representing a cross-correlation effect can be considered as a result of the Doppler shift as we demonstrate here.
Taking a derivative of $\Ev_{\rm tot}$ with respect to time, we have, using $\frac{\partial \Bv}{\partial t} =-\bm{\nabla}\times \Ev$,
\begin{align}
\frac{\partial \Ev_{\rm tot}}{\partial t} =\frac{\partial \Ev}{\partial t}+\frac{1}{\ez}[(\uv\cdot\bm{\nabla})\Ev-\bm{\nabla}(\uv\cdot\Ev)].
\end{align}
For plane waves with the wave vector $\kv\perp\Ev$, the last term in the above equation is orthogonal to the field $\Ev$; hence, it is neglected as was done for the linear effects in $\uv$. The time derivative is thus replaced by a \textquotedblleft covariant\textquotedblright\ one,
\begin{align}
 D_t\equiv \frac{\partial}{\partial t}+\frac{1}{\ez}(\uv\cdot\bm{\nabla}),
\end{align}
which is expected for a flowing medium \cite{Leonhardt99}. 
Therefore, the electromagnetic cross-correlation effect shown in Eq. (\ref{EBcross}) represents the Doppler shift because of a medium flow with velocity $\uv$. 
\\
\indent
It needs to be understood which type of intrinsic flow causes the Doppler shift of the electromagnetic field in solids.
A charge flow induces Joule heating; hence, it does not occur spontaneously in macroscopic solids. In contrast, an equilibrium flow of spin (spin current) can occur even at macroscopic scales, as known in magnets with noncollinear magnetization structures \cite{TKS_PR08,Phuc15}. 
Spin current can be spontaneously generated by the spin-orbit interaction because the spin current breaks the spatial-inversion symmetry but not the time-reversal symmetry. 
In fact, a recent study showed that such an intrinsic spin current generates the Dzyaloshinskii--Moriya interaction in magnetic materials as a result of the Doppler shift \cite{Kikuchi16}. 
The aim of this study is to demonstrate that the magnetic Rashba conductor induces the vector-coupling term of Eq. (\ref{ExB}), and the vector $\uv$ is given by the Rashba-induced spin gauge field $\bm{{\cal A}}_{\rm R}$. 
This result indicates that wave propagation is affected by the Doppler shift induced by the intrinsic spin current generated by the spin gauge field.
This is an interesting result as it indicates that an effective vector potential for an electron spin acts as an effective vector potential for an electromagnetic wave (light) as suggested by Refs. \cite{Sawada05,Leonhardt99}. 
The directional dichroism predicted in the magnetic Rashba conductor in Ref. \cite{Shibata16} can thus be explained based on this result. 
We should keep in mind, however, that this scenario is justified up to the linear order of $\uv$ representing the intrinsic flow in medium. 

\section{Cross-correlation effects on electric permittivity} \label{III}
The sections until now describe the effects originating from the vector coupling of Eq. (\ref{ExB}); however, there may be other effects induced by the quadrupole moment of the system. These effects are described in the section below. 
The cross-correlation effect because of the quadrupole moment is expressed as \cite{Spaldin08} 
\begin{align}
{\cal{H}}_{Q}&=\sum_{ij}Q_{ij}E_{i}B_{j} \label{qEB},
\end{align}
where $Q_{ij}$(=$Q_{ji}$) is the traceless quadrupole moment. 
The electric and magnetic fields representing the cross-correlation effects due to the total Hamiltonian 
${\cal{H}}_{u}+{\cal{H}}_{Q}$ after considering the vector $\bm{u}$ and the quadrupole moment $Q_{ij}$ are
\begin{align}
 \Ev_{\rm tot} =& \Ev+\frac{1}{\ez}\bm{P}, \nnr
 \Bv_{\rm tot}= & \Bv+\muz\Mv \label{qEBcross},
\end{align}
where $P_{\mu}\equiv(\uv\times\Bv)_{\mu}-\sum_{\nu}Q_{\mu\nu}B_{\nu}$ and $M_{\mu}\equiv(\uv\times\Ev)_{\mu}+\sum_{\nu}Q_{\mu\nu}E_{\nu}$ indicate effective electric polarization and effective magnetization, respectively. 
The electric permittivity tensor read from Eq. (\ref{qEBcross}) is 
\begin{align}
\epsilon_{ij}&=\epsilon^{(0)}_{ij}
-\frac{1}{\ez\omega}\sum_{l}
\biggl[\sum_{\nu}(Q_{i\nu}\epsilon_{\nu jl}+\epsilon_{\nu il}Q_{j\nu}) \nnr
&+(u_{i}\delta_{lj}+\delta_{il}u_{j})\biggl]k_{l}
+\frac{2}{\ez\omega}\delta_{ij}(\bm{u}\cdot\kv) \label{epqu},
\end{align}
where $\epsilon_{ijk}$ is a totally antisymmetric tensor, $\kv$ and $\omega$ are the wave vector and frequency of the electromagnetic field, respectively, and $\epsilon^{(0)}_{ij}$ is the contribution even in $k$. 
It can be seen that the cross-correlation effects by $\bm{u}$ and $Q_{ij}$ lead to generation of linear components in the wave vector $\kv$ that change signs by time-reversal \cite{Baranova77}. 
The second term on the right-hand side of Eq. (\ref{epqu}) indicates that the quadrupole moment contributes to the off-diagonal component of the electric permittivity tensor which affects only circularly-polarized waves. 
The optical responses was discussed in the context of correlation functions and it was shown that the quadrupole moment results in an optical activity such as the Faraday effect for polarized waves \cite{Nakano69}. 
The optical effect arising from the terms proportional to $u_{i}\delta_{lj}$ and $\delta_{il}u_{j}$ corresponds to an inverse Faraday effect induced by a spin-orbit interaction \cite{Mondal15}. 
By contrast, the third term on the right-hand side of Eq. (\ref{epqu}) means that the Doppler shift, $\uv\cdot\kv$, leads to the generation of the diagonal component which affects unpolarized waves. 
This Doppler shift term is known as magnetochiral dichroism, where magnetization $\bm{M}$ in chiral materials acts as the vector $\bm{u}$ \cite{Train08}. 
In the case of Rashba spin-orbit system, the Rashba field $\alphaRv$, which is invariant under time reversal, cannot induce $\bm{u}$ on its own. Our result indicates that the combination $\alphaRv\times \Mv$ plays the role of vector $\bm{u}$, suggesting possible applications of natural magnetic spin-orbit systems to novel optical materials. 

\section{Derivation of effective Hamiltonian}
In this section, we derive an effective Hamiltonian based on an imaginary-time path-integral formalism \cite{Sakita85} and confirm that the $\Ev$-$\Bv$ coupling in the magnetic Rashba conductor is described by ${\cal{H}}_{u}$ and ${\cal{H}}_{Q}$. 
Here, $\hbar=1$ is used for simplicity, where $\hbar$ is the Planck constant divided by $2\pi$. 
The system we consider consists of electrons with Rashba interaction and also interaction with magnetization and electromagnetic fields. 
The electrons are represented using annihilation and creation fields having two spin components, defined on an imaginary time $\tau$, ${c} (\bm{r},\tau)$ and $\bar{c} (\bm{r},\tau)$. 
The Hamiltonian is ${H} = {H}_{\text{0}}  +{H}_{sd} +{H}_{\rm R} + {H}_{\text{em}}$, where 
\begin{align} 
{H}_{\text{0}}  &=  \int {\rm{d}}^{3} r \left(\frac{1} {2m} |\bm{\nabla}{c}|^2 - \mu \bar{c} c \right)
\end{align}
describes the kinetic energy of conduction electrons measured from the Fermi energy. $m$ is the electron mass and $\mu$ is the chemical potential of the system. 
The second term, 
\begin{align} 
{H}_{sd} &= -J_{sd} \int {\rm{d}}^{3} r   \bm{M} \cdot \left(\bar{c} \bm{\sigma} c\right),
\end{align}
is the exchange interaction between the magnetization and conduction-electron spin. $J_{sd}$ is its strength, $\bm{M}$ is the magnetization vector, and $\bm{\sigma}$ is the vector of Pauli matrices. The magnetization is treated as spatially uniform and static. 
We consider a weak $sd$-exchange interaction up to the linear order. $H_{\rm R}$ is the Rashba spin-orbit interaction, 
\begin{align} 
{H}_{\rm R} &= \frac{i}{2}\int {\rm{d}}^{3} r \alphaRv \cdot \bar{c} (\overleftrightarrow{\nabla} \times \bm{\sigma}) c,
\end{align}
where $\bar{c} \overleftrightarrow{\nabla} c $ $\equiv$ $\bar{c}\left(\bm{\nabla} c \right)-\left(\bm{\nabla} \bar{c} \right)c $. 
The term ${H}_{\text{em}}$ describes the interaction between the conduction electron and the electromagnetic field, described by a gauge field $\bm{A}$. 
Taking into account the term arising from the Rashba spin-orbit interaction, the interaction reads as, 
\begin{align}
{H}_{\text{em}} &= - \int {\rm{d}}^{3} r \bm{A}
\cdot \left(\frac{ie}{2m} \bar{c} \overleftrightarrow{\nabla} c - \frac{e^2}{2m}  \bm{A} \bar{c} c-e\bar{c}(\alphaRv \times \bm{\sigma})c\right) \label{Hem},
\end{align}
where $-e$ is the electron charge ($e > 0$). 
The current density derived from Eq. (\ref{Hem}) is 
\begin{align}
 j_{i}^{\rm tot} 
 =& -j_{i}-\frac{e^2}{m}A_{i}\bar{c}c-2e\sum_{jk}\epsilon_{ijk}\alpha_{{\rm R},j} s_{k},
\end{align}
where 
\begin{align}
  j_{i}(\bm{r},\tau)\equiv & -\frac{ie} {2m}\bar{c}(\bm{r},\tau)\overleftrightarrow{\nabla_{i}}c(\bm{r},\tau), \nnr
  s_{\alpha}(\bm{r},\tau)\equiv & \frac{1}{2} \bar{c}(\bm{r},\tau)\sigma_{\alpha}c(\bm{r},\tau),
\end{align}
are the bare electric current density and spin density, respectively. 
\\
\indent
The effective Hamiltonian for the electromagnetic field, $H_{\rm {eff}}$, is calculated by integrating the electrons in the partition function 
$\mathcal{Z}(\bm{A})= \int \mathcal{D}\bar{c}\mathcal{D}ce^{-\int_0^\beta d\tau L(\bar{c},c,\bm{A})}$ as $\int_{0}^{\beta}{\rm d}\tau H_{\rm {eff}}=-{\rm Tr}\ln \mathcal{Z}$, 
where $\mathcal{D}$ denotes the path integral, $L= \int {\rm{d}}^{3} r \bar{c} \partial_\tau c+H$ is the imaginary-time Lagrangian, and $\beta$ is the inverse temperature. 
The partition function is perturbatively calculated to the second order in the gauge field. 
The result, diagramatically shown in Fig. 1, is 
\begin{align}
\ln \mathcal{Z} &=
-\int_{0}^{\beta} {\rm{d}}\tau\int {\rm{d}}^3r \sum_{\mu\nu}A_{\mu}A_{\nu}\frac{e^2}{2m}n_{\rm e}(\bm{r},\tau)\delta_{\mu\nu} \nnr
&+\frac{1}{2}\int_{0}^{\beta} {\rm{d}}\tau\int_{0}^{\beta} {\rm{d}}\tau'\int {\rm{d}}^3r\int {\rm{d}}^3r' \sum_{\mu\nu}A_{\mu}A_{\nu} \nonumber\\
&\times \biggl[
\chi_{jj}^{\mu\nu}(\bm{r},\bm{r}',\tau,\tau') +\chi_{sj}^{\mu\nu}(\bm{r},\bm{r}',\tau,\tau')+\chi_{js}^{\mu\nu}(\bm{r},\bm{r}',\tau,\tau') \nnr
&+\chi_{ss}^{\mu\nu}(\bm{r},\bm{r}',\tau,\tau') 
\biggl]. \label{lnZ}
\end{align}
Here, $n_{\rm e}(\bm{r},\tau)\equiv \langle\bar{c}(\bm{r},\tau)c(\bm{r},\tau)\rangle$ is the electron density, the thermal average $\langle\ \rangle$ is calculated in the equilibrium state determined by $\int_0^\beta d\tau L_{\text{0}}\equiv\int_0^\beta d\tau[ \int {\rm{d}}^{3} r \bar{c} \partial_\tau c+{H}_{\text{0}}  +{H}_{sd} +{H}_{\rm R}]$ and 
\begin{align}
\chi_{jj}^{\mu\nu}(\bm{r},\bm{r}',\tau,\tau')&\equiv \langle j_{\mu}(\bm{r},\tau)  j_{\nu}(\bm{r}',\tau') \rangle, \nonumber\\
\chi_{sj}^{\mu\nu}(\bm{r},\bm{r}',\tau,\tau')&\equiv 2e\sum_{m\alpha}\epsilon_{\mu m\alpha}\alpha_{{\rm R},m}\langle  s_{\alpha}(\bm{r},\tau) j_{\nu}(\bm{r}',\tau') \rangle,  \nonumber\\
\chi_{js}^{\mu\nu}(\bm{r},\bm{r}',\tau,\tau')&\equiv 2e\sum_{m\alpha}\epsilon_{\nu m\alpha}\alpha_{{\rm R},m}\langle  j_{\mu}(\bm{r},\tau) s_{\alpha}(\bm{r}',\tau')  \rangle, \nonumber\\
\chi_{ss}^{\mu\nu}(\bm{r},\bm{r}',\tau,\tau') &\equiv 4e^2\sum_{mo\alpha\beta}\epsilon_{\mu o\alpha}\epsilon_{\nu m\beta}\alpha_{{\rm R},o}\alpha_{{\rm R},m} \nnr
&\times\langle s_{\alpha}(\bm{r},\tau)s_{\beta}(\bm{r}',\tau') \rangle
\end{align}
represent the correlation functions of the current and the spin density. 
\begin{figure}
\centering
\includegraphics[width=8.6cm]{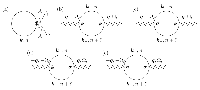}
\caption{The Feynman diagrams for the effective Hamiltonian. Solid lines represent the conducting electrons' Green's function and the wavy lines denote the gauge field, respectively. Diagrams, $(\rm a)$, $(\rm b)$, $(\rm c)$, $(\rm d)$, and $(\rm e)$ correspond to the contributions of $n_{\rm e}$, $\chi_{jj}^{\mu\nu}$, $\chi_{sj}^{\mu\nu}$, $\chi_{js}^{\mu\nu}$, and $\chi_{ss}^{\mu\nu}$ in Eq. (\ref{lnZ}), respectively.}
\end{figure}
The electron density $n_{\rm e}$ is expressed as $n_{\rm e}=-\frac{1}{\beta V}\sum_{n,\kv}{\rm tr}[{\mathscr{G}}_{\kv,n,\bm{M}}]$, where ${\rm tr}$ is the trace over spin space, $V$ is the volume of the system, and 
\begin{align}
{\mathscr{G}}_{\kv,n,\bm{M}}&\equiv 
\frac{1}{i\omega_{n}-\epsilon_{\kv}-\bm{\gamma}_{\kv,\bm{M}}\cdot\bm{\sigma}+i\eta{\rm sgn}(n)}
\end{align}
is the thermal Green's function for electrons that includes the Rashba and $sd$-exchange interactions. 
Here, $\kv$ and $\omega_n\equiv\frac{(2n+1)\pi}{\beta}$ ( $n$ is an integer) indicate the wave vector and fermionic thermal frequency, respectively, $\epsilon_{\kv}=\frac{k^2}{2m}-\mu$ is the electron energy measured from the Fermi energy, and $\bm{\gamma}_{\kv,\bm{M}}\equiv \bm{\gamma}_{\kv}-J_{sd}\bm{M}$ with  $\bm{\gamma}_{\kv}\equiv \kv\times \alphaRv$. 
We have included a finite electron-elastic-scattering lifetime $\tau_{\rm e}$ as an imaginary part, $\eta\equiv \frac{1}{2\tau_{\rm e}}$, and ${\rm sgn}(n)$ $\equiv$ $1$ and $-1$ for $n > 0$ and $n < 0$, respectively. 
In terms of the Green's function, correlation functions read as
\begin{align}
\chi_{jj}^{\mu\nu}(\qv,i\Omega_{\ell},\bm{M})&=-\frac{e^2}{m^2\beta V}\sum_{n,\kv}k_{\mu}k_{\nu} {\rm tr} [{\mathscr{G}}_{\kv_{+},n+\ell,\bm{M}}{\mathscr{G}}_{\kv_{-},n,\bm{M}}] ,\nonumber\\
\chi_{sj}^{\mu\nu}(\qv,i\Omega_{\ell},\bm{M})&=-\frac{e^2}{m\beta V}\sum_{n,\kv}\sum_{m\alpha}\epsilon_{\mu m\alpha}\alpha_{{\rm R},m}k_{\nu} \nnr
&\times{\rm tr} [\sigma_{\alpha}{\mathscr{G}}_{\kv_{+},n+\ell,\bm{M}}{\mathscr{G}}_{\kv_{-},n,\bm{M}}] ,\nonumber\\
\chi_{js}^{\mu\nu}(\qv,i\Omega_{\ell},\bm{M})&=-\frac{e^2}{m\beta V}\sum_{n,\kv}\sum_{m\alpha}\epsilon_{\nu m\alpha}\alpha_{{\rm R},m}k_{\mu} \nnr
&\times{\rm tr} [{\mathscr{G}}_{\kv_{+},n+\ell,\bm{M}}\sigma_{\alpha}{\mathscr{G}}_{\kv_{-},n,\bm{M}}] ,\nonumber\\
\chi_{ss}^{\mu\nu}(\qv,i\Omega_{\ell},\bm{M})&=-\frac{e^2}{\beta V}\sum_{n,\kv}\sum_{mo\alpha\beta}\epsilon_{\mu o\alpha}\epsilon_{\nu m\beta}\alpha_{{\rm R},o}\alpha_{{\rm R},m} \nnr
&\times{\rm tr} [\sigma_{\alpha}{\mathscr{G}}_{\kv_{+},n+\ell,\bm{M}}\sigma_{\beta}{\mathscr{G}}_{\kv_{-},n,\bm{M}}]  \label{chi},
\end{align}
where $\kv_{\pm}\equiv\kv\pm\frac{\qv}{2}$. 
The wave vector and thermal frequency carried by the gauge field are denoted by $\qv$ and $\Omega_{\ell}$, respectively.
($\Omega_\ell$ $\equiv$ $\frac{2\pi \ell}{\beta}$ is a bosonic thermal frequency). 
\\
\indent
We are interested in a low-energy long-wavelength effective Hamiltonian; hence, we expand the correlation functions given in Eq. (\ref{chi}) with respect to $\qv$. 
The result up to the linear order of ${\qv}$ and $\Mv$ is (see Appendix \ref{A}) 
\begin{align}
\chi_{\mu\nu}^{(1)}(\qv,i\Omega_{\ell},\bm{M})
&\equiv \chi_{jj}^{\mu\nu}+\chi_{sj}^{\mu\nu}+\chi_{js}^{\mu\nu}+\chi_{ss}^{\mu\nu} \nnr
&=
g_1(i\Omega_{\ell}) (\bm{{\cal A}}_{\rm R}\cdot \qv)\delta_{\mu\nu}^{\perp} \nnr
&+g_2(i\Omega_{\ell}) ({\cal A}_{{\rm R},\mu}q_{\nu}+q_{\mu}{\cal A}_{{\rm R},\nu}) \nnr
&+g_3(i\Omega_{\ell}) [M_{\mu}^{\perp}(\bm{\alpha}_{\rm R}\times\qv)_{\nu}+(\bm{\alpha}_{\rm R}\times\qv)_{\mu}M_{\nu}^{\perp}]
 \label{xtot},
\end{align}
where $\delta_{\mu\nu}^{\perp} \equiv\delta_{\mu\nu}-\hat{\alpha}_{{\rm R},\mu}\hat{\alpha}_{{\rm R},\nu}$, $\hat{\bm{\alpha}}_{\rm R}\equiv\alphaRv/|\alphaRv|$ is a unit vector representing the direction of the Rashba field, and $\Mv^{\perp}\equiv \Mv-(\hat{\bm{\alpha}}_{\rm R}\cdot \Mv)\hat{\bm{\alpha}}_{\rm R}$.
We have considered the case where $\alphaRv\cdot\qv=0$ to simplify the angular integration calculation with respect to $\kv$. This expression otherwise consists of more terms. 
The vector $\bm{{\cal A}}_{\rm R}\equiv\bm{\alpha}_{\rm R}\times\Mv$ represents an effective spin gauge field and a toroidal moment. 
The coefficients $g_{1}$, $g_{2}$, and $g_{3}$ are defined in Appendix. \ref{B}.
\\
\indent
Equation (\ref{xtot}) contributes to directional dichroism as it is linear in $\qv$. 
The first and second terms of this equation proportional to $g_1$ and $g_2$ indicate that the existence of the toroidal moment governs the anomalous light propagation, whereas the terms proportional to $M_{\mu}^{\perp}(\bm{\alpha}_{\rm R}\times\qv)_{\nu}$ and $(\bm{\alpha}_{\rm R}\times\qv)_{\mu}M_{\nu}^{\perp}$ represent the contributions arising from the effective quadrupole moment, as shown in the effective theory \cite{Shibata16}. 
The result can be simplified using the gauge invariance of the effective Hamiltonian (see Appendix \ref{C}).
It was observed that $g_{1}=-2g_{2}$ is restricted by the gauge invariance, whereas $g_{3}$ is not restricted by this invariance. 
\\
\indent
Carrying out the analytic continuation, the coefficient, $g_{1}$ is calculated. $g_{1}$ is written by rewriting the summation over the thermal frequency using the contour integral $(z\equiv i\omega_{n})$ as
\begin{align}
g_1(i\Omega_{\ell}) &=-2\frac{J_{sd}}{2^4}\left(\frac{e}{m}\right)^2\sum_{\kv}\sum_{\sigma_{1}\sigma_{2}\sigma_{3}\sigma_{4}}\left(\frac{\gamma_{\kv}}{\alpha_{\rm R}}\right)^2 
\biggl[\sigma_{1}\sigma_{2} \nnr
&+\frac{m\alpha_{\rm R}^2}{\gamma_{\kv}}(\sigma_{2}\sigma_{3}\sigma_{4}+2\sigma_{2})\biggl] 
\int_{C}\frac{{\rm d}z}{2\pi i}f(z) {\rm g}_{\kv,\sigma_{1}}(z){\rm g}_{\kv,\sigma_{2}}(z) \nnr
&\times[{\rm g}_{\kv,\sigma_{3}}(z+i\Omega_{\ell}){\rm g}_{\kv,\sigma_{4}}(z+i\Omega_{\ell}) \nnr
&-{\rm g}_{\kv,\sigma_{3}}(z-i\Omega_{\ell}){\rm g}_{\kv,\sigma_{4}}(z-i\Omega_{\ell})],
\end{align}
where $C$ is a counterclockwise contour surrounding the imaginary axis \cite{ADG75,AS06}, $\gamma_{\kv}\equiv|\bm{\gamma}_{\kv}|$, ${\rm g}_{\kv,\sigma}(z)\equiv \left[z-\epsilon_{\kv}^{\sigma}+i\eta{\rm sgn}({\rm Im} [z])\right]^{-1}$ with $\epsilon_{\kv}^{\sigma}=\epsilon_{\kv}+\sigma\gamma_{\kv}$ is the Green's function diagonalized in spin space, $\sigma_{i}$ = $\pm$ ($i=1\sim 4$) are the diagonalized spin indices, ${\rm Im}$ denotes the imaginary part, and $f(z)\equiv(e^{\beta z}+1)^{-1}$ is the Fermi--Dirac distribution function. We expand the coefficient $g_{1}$ with respect to external frequency $\Omega$ after the analytic continuation to $\Omega +i0\equiv i\Omega_{\ell}$ \cite{ADG75,AS06}, where $i0$ denotes a small imaginary part. The result up to the linear order in $\Omega$ is
\begin{align}
g_1(\Omega)&= 2\Omega g,
\end{align}
with
\begin{align}
g&\equiv i\frac{J_{sd}}{2^4}\left(\frac{e}{m}\right)^2\sum_{\kv,\omega}\sum_{\sigma_{1}\sigma_{2}\sigma_{3}\sigma_{4}}\left(\frac{\gamma_{\kv}}{\alpha_{\rm R}}\right)^2 \nnr
&\times\biggl[\sigma_{1}\sigma_{2}+\frac{m\alpha_{\rm R}^2}{\gamma_{\kv}}(\sigma_{2}\sigma_{3}\sigma_{4}+2\sigma_{2})\biggl] \nnr
&\times
\begin{Bmatrix}
f_{\omega}[({\rm g}^{\rm r}_{\kv,\omega,\sigma_{1}}{\rm g}^{\rm r}_{\kv,\omega,\sigma_{2}})\overleftrightarrow{\partial_{\omega}}({\rm g}^{\rm r}_{\kv,\omega,\sigma_{3}}{\rm g}^{\rm r}_{\kv,\omega,\sigma_{4}})-c.c.]
\\
+f_{\omega}'[ {\rm g}^{\rm a}_{\kv,\omega,\sigma_{1}}{\rm g}^{\rm a}_{\kv,\omega,\sigma_{2}}{\rm g}^{\rm r}_{\kv,\omega,\sigma_{3}}{\rm g}^{\rm r}_{\kv,\omega,\sigma_{4}}-c.c.]
\end{Bmatrix},
\end{align}
where ${\rm g}_{\kv,\omega,\sigma}^{\rm r}\equiv(\omega-\epsilon_{\kv}^{\sigma}+i\eta)^{-1}$ and ${\rm g}_{\kv,\omega,\sigma}^{\rm a}=({\rm g}_{\kv,\omega,\sigma}^{\rm r})^{\ast}$ are the retarded and advanced Green's functions of the conduction electron with wave vector $\kv$ and angular frequency $\omega$, respectively. $\sum_{\omega}\equiv\int_{-\infty}^{\infty}\frac{{\rm{d}}\omega}{2\pi}$, $\partial_{\omega}\equiv\frac{\partial}{\partial\omega}$, $f_{\omega}$=$(e^{\beta \omega}+1)^{-1}$, and $f_{\omega}'\equiv\partial_{\omega}f_{\omega}$. 
$g_{3}$ is also given by
\begin{align}
g_3(\Omega)&= \Omega \lambda ,
\end{align}
with
\begin{align}
\lambda&\equiv\frac{i}{2}\frac{J_{sd}}{2^4}\left(\frac{e}{m}\right)^2\sum_{\kv,\omega}\sum_{\sigma_{1}\sigma_{2}\sigma_{3}\sigma_{4}}\sigma_{1}\sigma_{2}\left(\frac{\gamma_{\kv}}{\alpha_{\rm R}}\right)^2 \nnr
&\times
\begin{Bmatrix}
f_{\omega}[({\rm g}^{\rm r}_{\kv,\omega,\sigma_{1}}{\rm g}^{\rm r}_{\kv,\omega,\sigma_{2}})\overleftrightarrow{\partial_{\omega}}({\rm g}^{\rm r}_{\kv,\omega,\sigma_{3}}{\rm g}^{\rm r}_{\kv,\omega,\sigma_{4}})-c.c.]
\\
+f_{\omega}'[ {\rm g}^{\rm a}_{\kv,\omega,\sigma_{1}}{\rm g}^{\rm a}_{\kv,\omega,\sigma_{2}}{\rm g}^{\rm r}_{\kv,\omega,\sigma_{3}}{\rm g}^{\rm r}_{\kv,\omega,\sigma_{4}}-c.c.]
\end{Bmatrix}.
\end{align}
The effective Hamiltonian describing the directional dichroism is finally obtained as 
\begin{align}
H_{\rm eff}&=H_{\cal{A}_{\rm R}} + H_{Q}, 
\end{align}
with
\begin{align}
H_{\cal{A}_{\rm R}}
&=g\int {\rm d}^3r \bm{{\cal A}}_{\rm R}\cdot (\Ev\times\Bv), \label{HAr} \\
H_{Q}
&=\lambda \int {\rm d}^3r \sum_{\mu\nu}Q_{\mu\nu}E_{\mu}B_{\nu}  \label{HQ},
\end{align}
where $Q_{\mu\nu}\equiv M_{\mu}^{\perp}\alpha_{{\rm R},\nu}$ is the effective quadrupole moment and $\Ev\equiv -\dot{\Av}$ and $\Bv\equiv\bm{\nabla}\times\Av$ are the electric and magnetic fields, respectively. 
\\
\indent
The effective Hamiltonian of Secs. \ref{II} and \ref{III}, ${\cal{H}}_{u}+{\cal{H}}_{Q}$ with $\bm{u}=\bm{{\cal A}}_{\rm R}$ is thus justified by microscopic derivation based on the imaginary-time path-integral formalism. 
The toroidal moment thus leads to the Doppler shift and this is the origin of the magnetic Rashba conductor that exhibits directional dichroism discussed in Sec. \ref{II}. 
\\
\indent
The coupling constant $g$ consists of terms proportional to $f_{\omega}$ and $f_{\omega}'$ but it cannot be expressed only in terms of $f_{\omega}'$ (Fermi surface term) using integration by parts. 
The $f_{\omega}'$ term represents the contribution arising from the state of electrons in the vicinity of a Fermi surface. 
The contribution of this term is finite in a metallic system, whereas it does not exist for insulators such as multiferroics. 
In contrast, the $f_{\omega}$ term (Fermi sea term) arises from the equilibrium property of the electrons forming the Fermi sea. The contribution of this term is finite in insulator systems; therefore,  directional dichroism can be induced even in broad materials. 

\section{Optical response of Weyl semimetal} \label{V}
So far, we have considered the electromagnetic cross-correlation effects in a nonrelativistic Rashba spin-orbit system. The cross-correlation effects also occur in a relativistic spin-orbit system such as 3+1-dimensional Weyl semimetal. 
It is known that the Weyl semimetal realized in multilayer of topological insulator and magnetic insulator (Ref. \cite{Burkov11}) induces topological electromagnetic response such as the anomalous Hall effect and chiral magnetic effect \cite{Vazifeh13}. 
\\
\indent
A Hamiltonian representing the Weyl spin-orbit system is made up of two terms, 
one describing the diagonal components and the other denoting the off-diagonal components (mass term) 
as \cite{Zyuzin12, Goswami13}
\begin{align}
{\cal{H}_{\rm Weyl}}=\tau^{z}(\bm{\sigma}\cdot\bm{k})+\tau^{z} b_{0}+\bm{\sigma}\cdot\bm{b},
\end{align}
where the first term is the Dirac Hamiltonian being linear in the wave vector $\kv$ of the Dirac electrons, $\bm{\tau}$ is the Weyl node degree of freedom, $\bm{\sigma}$ is the conduction-valence band degree of freedom, 
$b_{0}$ is a constant reflecting the breaking of spatial-inversion symmetry, and $\bm{b}$ is a constant vector reflecting the breaking of time-reversal invariance.
$\cal{H}_{\rm Weyl}$ is $4\times4$ Hamiltonian because of having the Weyl nodes of opposite chirality and conduction-valence band degrees of freedom. By the Fujikawa's method \cite{Fujikawa04}, the effective Hamioltonian describing the electromagnetic response of the Weyl system is shown to be \cite{Zyuzin12, Goswami13}
\begin{align}
H_{\theta}&=\frac{\alpha}{8\pi}\sum_{\mu\nu\alpha\beta}\int {\rm d}^3r \theta(\bm{r},t)\epsilon^{\mu\nu\alpha\beta}F_{\mu\nu}F_{\alpha\beta} , \label{Hthetaw}
\end{align}
where we set $\hbar=c=\epsilon_{0}=1$, $\alpha\equiv\frac{e^2}{4\pi}$ is the fine structure constant, and $F_{\mu\nu}\equiv\partial_{\mu}A_{\nu}-\partial_{\nu}A_{\mu}$ is the field strength of the electromagnetic field. $\theta(\bm{r},t)\equiv 2(\bm{b}\cdot\bm{r}-b_{0}t)$ is the field depending linearly on space $\bm{r}$ and time $t$. 
The total electric and magnetic fields representing the cross-correlation effects due to the Hamiltonian (\ref{Hthetaw}) become 
$\Ev_{\rm tot} = \Ev+\bm{P}$ and $\Bv_{\rm tot}= \Bv+\Mv$ 
with 
\begin{align}
\bm{P} &\equiv -\frac{\alpha}{\pi}\theta\Bv , \nnr
\Mv &\equiv\frac{\alpha}{\pi}\theta\Ev .
\end{align}
By a similar discussion as in Sec. \ref{III}, we see that the field $\theta$ leads to the off-diagonal component of the electric permittivity tensor as 
\begin{align}
\epsilon_{\mu\nu}&=\delta_{\mu\nu}
+\frac{2\alpha}{\pi\omega}\sum_{l}\epsilon_{\mu\nu l}(b_{0}k_{l}+b_{l}\omega)
\label{etheta} ,
\end{align}
where $\kv$ and $\omega$ are the wave vector and frequency of the electromagnetic field. 
Equation (\ref{etheta}) indicates that the field $\theta$ induces the Faraday effect for circularly polarized waves \cite{Kargarian15}. 
Comparing Eq. (\ref{epqu}) with Eq. (\ref{etheta}), the field $\theta$ turns out to play the same role as the quadrupole moment in the context of the magneto-optical response. 
The Doppler shift term does not appear in the Weyl system, as the system is Lorentz invariant. 
\\

\section{Summary}
We have derived an effective Hamiltonian describing directional dichroism in a magnetic Rashba conductor, and showed that it is 
made up of a toroidal-moment term and a quadrupole-moment term, as in insulator multiferroics. 
The toroidal-moment term is given by the vector coupling between the toroidal moment and the Poynting vector, such that this term leads to directional dichroism as a result of the Doppler shift, whereas the quadrupole-moment term induces a magneto-optical phenomenon such as the Faraday effect when circularly-polarized waves are applied. 
The microscopic analysis done in this study indicates that the toroidal moment plays the role of an effective vector potential for light, causing dichroism as a result of the Doppler shift. 
The effective Hamiltonian approach clearly shows that electromagnetic cross-correlation effects in the magnetic Rashba system is qualitatively distinct from those in relativistic Weyl systems described by the topological $\theta$ term.


\acknowledgements
The authors thank T. Kikuchi, A. Shitade, and H. Kohno for valuable discussions.
This work was supported by a Grant-in-Aid for Scientific Research (C) (Grant No. 25400344) from Japan Society for the Promotion of Science and Grant-in-Aid for Scientific Research on Innovative Areas (Grant No.26103006) from The Ministry of Education, Culture, Sports, Science, and Technology (MEXT), Japan. One of the authors (H. K.) was supported by RIKEN Junior Associate Program.




\appendix
\begin{widetext}
\section{Derivation of Eq. (\ref{xtot})} \label{A}
This section shows the detailed derivation of Eq. (\ref{xtot}) from Eq. (\ref{chi}). 
Performing the trace over the spin using 
${\rm tr}[\sigma_{\alpha}\sigma_{\beta}\sigma_{\gamma}\sigma_{\delta}]=2(\delta_{\alpha\beta}\delta_{\gamma\delta}
+\delta_{\beta\gamma}\delta_{\alpha\delta}-\delta_{\alpha\gamma}\delta_{\beta\delta})$, 
the result up to the linear order in $\qv$ and $\Mv$ is obtained as
\begin{align}
\chi_{jj}^{\mu\nu}(\qv,i\Omega_{\ell},\bm{M})&\simeq -\frac{e^2J_{sd}}{m^2\beta V}\sum_{n,\kv}\sum_{\alpha}(-M_{\alpha})(k_{\mu}q_{\nu}+q_{\mu}k_{\nu})\hat{\gamma}_{\kv}^{\alpha}
\begin{Bmatrix} 2h_{\kv,n}j_{\kv,n}(h_{\kv,n+\ell}-h_{\kv,n-\ell})  \\ +(h_{\kv,n}^2+j_{\kv,n}^2)(j_{\kv,n+\ell}-j_{\kv,n-\ell}) \end{Bmatrix} ,
\nonumber\\
\chi_{sj}^{\mu\nu}(\qv,i\Omega_{\ell},\bm{M})&\simeq  -\frac{e^2J_{sd}}{m\beta V}\sum_{n,\kv}\sum_{m\alpha\beta\rho}\epsilon_{\mu m\alpha}\alpha_{{\rm R},m}(-M_{\beta}) \nonumber\\
&\times 
\begin{Bmatrix}
-\frac{4}{m}q_{\rho} (k_{\nu}k_{\rho}\delta_{\alpha\beta}-k_{\nu}k_{\rho}\hat{\gamma}_{\kv}^{\alpha}\hat{\gamma}_{\kv}^{\beta})j_{\kv,n}^2(h_{\kv,n+\ell}^2-h_{\kv,n-\ell}^2)
\\
+4\gamma_{\qv}^{\rho}(\delta_{\rho\alpha}k_{\nu}\hat{\gamma}_{\kv}^{\beta}-\delta_{\alpha\beta}k_{\nu}\hat{\gamma}_{\kv}^{\rho})h_{\kv,n}j_{\kv,n}[(h_{\kv,n+\ell}^2-h_{\kv,n-\ell}^2)-(j_{\kv,n+\ell}^2-j_{\kv,n-\ell}^2)]
\\
+q_{\rho}\delta_{\rho j} 
\begin{bmatrix}
\delta_{\alpha\beta}(h_{\kv,n}^2-j_{\kv,n}^2)(h_{\kv,n+\ell}-h_{\kv,n-\ell}) \\ 
+2\hat{\gamma}_{\kv}^{\alpha}\hat{\gamma}_{\kv}^{\beta}[h_{\kv,n}j_{\kv,n}(j_{\kv,n+\ell}-j_{\kv,n-\ell})+j_{\kv,n}^2(h_{\kv,n+\ell}-h_{\kv,n-\ell})]
\end{bmatrix}
\end{Bmatrix},
\nonumber\\
\chi_{ss}^{\mu\nu}(\qv,i\Omega_{\ell},\bm{M})&\simeq  -\frac{e^2J_{sd}}{\beta V}\sum_{n,\kv}\sum_{mo\alpha\beta\gamma}\sum_{\mu'\nu'\rho}\epsilon_{\mu o\alpha}\epsilon_{\nu m\beta}\alpha_{{\rm R},o}\alpha_{{\rm R},m}(-M_{\gamma}) \nonumber\\
&\times
\begin{Bmatrix}
-\frac{4}{m}q_{\nu'}
\begin{bmatrix}
k_{\nu'}\hat{\gamma}_{\kv}^{\mu'}(\epsilon_{\alpha \mu' \rho}\epsilon_{\rho \gamma \beta}+\epsilon_{\alpha \gamma \rho}\epsilon_{\rho \mu' \beta})
h_{\kv,n}j_{\kv,n}[(h_{\kv,n+\ell}^2-h_{\kv,n-\ell}^2)+(j_{\kv,n+\ell}^2-j_{\kv,n-\ell}^2)] \\
+2{\displaystyle \sum_{ijkp}}\epsilon_{\gamma i j}\epsilon_{j k p}k_{\nu'}\hat{\gamma}_{\kv}^{i}\hat{\gamma}_{\kv}^{k}\hat{\gamma}_{\kv}^{\mu'}(\epsilon_{\alpha p \rho}\epsilon_{\rho \mu' \beta}+\delta_{\mu' \beta}
\delta_{\alpha p})h_{\kv,n}j_{\kv,n}(j_{\kv,n+\ell}^2-j_{\kv,n-\ell}^2)
\end{bmatrix}
\\
+4{\displaystyle \sum_{ijk}}\gamma_{\qv}^{\mu'}\hat{\gamma}_{\kv}^{i}\hat{\gamma}_{\kv}^{k}\epsilon_{jk\nu'}
\begin{bmatrix}
\epsilon_{\mu' ij}
(\epsilon_{\alpha \gamma \rho}\epsilon_{\rho \nu' \beta}+\delta_{\nu' \beta}\delta_{\alpha \gamma}) \\
-\epsilon_{\gamma i j}
(\epsilon_{\alpha \nu' \rho}\epsilon_{\rho \mu' \beta}+\delta_{\mu' \beta}\delta_{\alpha \nu'})
\end{bmatrix}
h_{\kv,n}^2(j_{\kv,n+\ell}^2-j_{\kv,n-\ell}^2)
\end{Bmatrix} ,
\end{align}
where $\bm{\hat{\gamma}}_{\kv}\equiv \bm{\gamma}_{\kv}/|\bm{\gamma}_{\kv}|$, $\bm{\gamma}_{\qv}\equiv\qv\times\alphaRv$, and $h_{\kv,n}$ and $j_{\kv,n}$ are defined as 
\begin{align}
h_{\kv,n}&\equiv\frac{1}{2}\sum_{\sigma=\pm}{\rm g}_{\kv,n,\sigma} ,\nnr 
j_{\kv,n}&\equiv\frac{1}{2}\sum_{\sigma=\pm}\sigma {\rm g}_{\kv,n,\sigma} \label{hj},
\end{align}
where ${\rm g}_{\kv,n,\sigma}\equiv \left[i\omega_{n}-\epsilon_{\kv}^{\sigma}+i\eta{\rm sgn}(n)\right]^{-1}$ with  $\epsilon_{\kv}^{\sigma}=\epsilon_{\kv}+\sigma\gamma_{\kv}$ ($\gamma_{\kv}\equiv|\bm{\gamma}_{\kv}|$) is the Green's function diagonalized in the spin space and $\sigma$ = $\pm$ is the diagonalized spin index. 
\\
\indent
To carry out the $\kv$ integral, we choose the $z$ axis along the Rashba field, i.e., $\bm{\alpha}_{\rm R}=\alpha_{\rm R}\hat{\bm{z}}$ ($\hat{\bm{z}}\equiv(0,0,1)$) and $\kv$ is represented using the polar and 
azimuthal angles $\theta$ and $\varphi$, respectively. 
We consider an electromagnetic wave with $q_{z}=0$, i.e., propagation perpendicular to the Rashba field. 
Let us first calculate $\chi_{jj}^{\mu\nu}$. Calculating the integral over $\varphi$, 
we obtain 
\begin{align}
\chi_{jj}^{\mu\nu}(\qv,i\Omega_{\ell},\bm{M})&= -\frac{J_{sd}}{\beta}\left(\frac{e}{m}\right)^2\sum_{n,\kv}\left(-\frac{\gamma_{\kv}}{2\alpha_{\rm R}^2}\right)\phi_{\kv,n}^{(0)}(i\Omega_{\ell})({\cal A}_{{\rm R},\mu}q_{\nu}+q_{\mu}{\cal A}_{{\rm R},\nu}) \label{xjj},
\end{align}
with 
\begin{align}
\phi_{\kv,n}^{(0)}(i\Omega_{\ell})\equiv 2h_{\kv,n}j_{\kv,n}(h_{\kv,n+\ell}-h_{\kv,n-\ell})+(h_{\kv,n}^2+j_{\kv,n}^2)(j_{\kv,n+\ell}-j_{\kv,n-\ell}) \label{phi0}.
\end{align}
Here, $\bm{{\cal A}}_{\rm R}\equiv\bm{\alpha}_{\rm R}\times\Mv$. 
Other correlation functions are calculated similarly as
\begin{align}
\chi_{sj}^{\mu\nu}(\qv,i\Omega_{\ell},\bm{M})&= -\frac{J_{sd}}{\beta}\left(\frac{e}{m}\right)^2\sum_{n,\kv} 
\begin{Bmatrix}
-\frac{\gamma_{\kv}^2}{2\alpha_{\rm R}^2}\phi_{\kv,n}^{(1)}(i\Omega_{\ell})[2(\bm{{\cal A}}_{\rm R}\cdot \qv)\delta_{\mu\nu}^{\perp} -3{\cal A}_{{\rm R},\mu}q_{\nu}+M_{\mu}^{\perp}(\bm{\alpha}_{\rm R}\times\qv)_{\nu}] \\
+2m\alpha_{\rm R}\gamma_{\kv}\phi_{\kv,n}^{(2)}(i\Omega_{\ell})
({\cal A}_{{\rm R},\mu}q_{\nu}-q_{\mu}{\cal A}_{{\rm R},\nu}) -m\phi_{\kv,n}^{(3)}(i\Omega_{\ell}){\cal A}_{{\rm R},\mu}q_{\nu}
\end{Bmatrix},
\nonumber\\
\chi_{ss}^{\mu\nu}(\qv,i\Omega_{\ell},\bm{M})&= -\frac{J_{sd}}{\beta}\left(\frac{e}{m}\right)^2\sum_{n,\kv}(-2m\gamma_{\kv}) 
\begin{Bmatrix}
\phi_{\kv,n}^{(4)}(i\Omega_{\ell})(\bm{{\cal A}}_{\rm R}\cdot \qv)\delta_{\mu\nu}^{\perp} \\
+\phi_{\kv,n}^{(5)}(i\Omega_{\ell})[2(\bm{{\cal A}}_{\rm R}\cdot \qv)\delta_{\mu\nu}^{\perp}-({\cal A}_{{\rm R},\mu}q_{\nu}+q_{\mu}{\cal A}_{{\rm R},\nu})]
\end{Bmatrix}, \label{xsjss}
\end{align}
with
\begin{align}
\phi_{\kv,n}^{(1)}(i\Omega_{\ell})&\equiv j_{\kv,n}^2(h_{\kv,n+\ell}^2-h_{\kv,n-\ell}^2),
\nonumber\\
\phi_{\kv,n}^{(2)}(i\Omega_{\ell})&\equiv h_{\kv,n}j_{\kv,n}[(h_{\kv,n+\ell}^2-h_{\kv,n-\ell}^2)-(j_{\kv,n+\ell}^2-j_{\kv,n-\ell}^2)],
\nonumber\\
\phi_{\kv,n}^{(3)}(i\Omega_{\ell})&\equiv h_{\kv,n}^2(h_{\kv,n+\ell}-h_{\kv,n-\ell})+ h_{\kv,n}j_{\kv,n}(j_{\kv,n+\ell}-j_{\kv,n-\ell}),
\nonumber\\
\phi_{\kv,n}^{(4)}(i\Omega_{\ell})&\equiv  h_{\kv,n}j_{\kv,n}(j_{\kv,n+\ell}^2-j_{\kv,n-\ell}^2),
\nonumber\\
\phi_{\kv,n}^{(5)}(i\Omega_{\ell})&\equiv  h_{\kv,n}j_{\kv,n}(h_{\kv,n+\ell}^2-h_{\kv,n-\ell}^2)
\label{phi},
\end{align}
where $\delta_{\mu\nu}^{\perp} \equiv\delta_{\mu\nu}-\hat{\alpha}_{{\rm R},\mu}\hat{\alpha}_{{\rm R},\nu}$, $\hat{\bm{\alpha}}_{\rm R}\equiv\alphaRv/|\alphaRv|$, and $\Mv^{\perp}\equiv \Mv-(\hat{\bm{\alpha}}_{\rm R}\cdot \Mv)\hat{\bm{\alpha}}_{\rm R}$. 
\\
\indent
The results of Eqs. (\ref{xjj}) and (\ref{xsjss}) are summarized in Eq. (\ref{xtot}). The coefficients $g_{1}$, $g_{2}$, and $g_{3}$ are defined in Appendix \ref{B}.

\section{Definition of coefficients $g_{1}$, $g_{2}$, and $g_{3}$} \label{B}
From (\ref{xjj}) and (\ref{xsjss}), coefficients $g_{1}$, $g_{2}$, and $g_{3}$ of Eq. (\ref{xtot}) are obtained as
\begin{align}
g_1(i\Omega_{\ell}) &\equiv-\frac{J_{sd}}{\beta}\left(\frac{e}{m}\right)^2\sum_{n,\kv}
\left(-\frac{2\gamma_{\kv}^2}{\alpha_{\rm R}^2}\right)\biggl[\phi_{\kv,n}^{(1)}(i\Omega_{\ell})+\frac{m\alpha_{\rm R}^2}{\gamma_{\kv}}(\phi_{\kv,n}^{(4)}(i\Omega_{\ell})+2\phi_{\kv,n}^{(5)}(i\Omega_{\ell}))\biggl],
\nonumber\\
g_2(i\Omega_{\ell}) &\equiv-\frac{J_{sd}}{\beta}\left(\frac{e}{m}\right)^2\sum_{n,\kv}
\frac{\gamma_{\kv}^2}{2\alpha_{\rm R}^2}
\biggl[
-\frac{1}{\gamma_{\kv}}\phi_{\kv,n}^{(0)}(i\Omega_{\ell})+3\phi_{\kv,n}^{(1)}(i\Omega_{\ell})
-\frac{2m\alpha_{\rm R}^2}{\gamma_{\kv}^2}\phi_{\kv,n}^{(3)}(i\Omega_{\ell})
+\frac{4m\alpha_{\rm R}^2}{\gamma_{\kv}}\phi_{\kv,n}^{(5)}(i\Omega_{\ell})
\biggl],
\nonumber\\
g_3(i\Omega_{\ell}) &\equiv-\frac{J_{sd}}{\beta}\left(\frac{e}{m}\right)^2\sum_{n,\kv}
\left(-\frac{\gamma_{\kv}^2}{2\alpha_{\rm R}^2}\right)\phi_{\kv,n}^{(1)}(i\Omega_{\ell})
.
\end{align}
Using Eqs. (\ref{hj}) and (\ref{phi}), the coefficients $g_{1}$ and $g_3$ can be written as
\begin{align}
g_1(i\Omega_{\ell}) &= -2\frac{J_{sd}}{2^4}\left(\frac{e}{m}\right)^2\sum_{\kv}\sum_{\sigma_{1}\sigma_{2}\sigma_{3}\sigma_{4}}
\left(\frac{\gamma_{\kv}}{\alpha_{\rm R}}\right)^2\biggl[\sigma_{1}\sigma_{2}+\frac{m\alpha_{\rm R}^2}{\gamma_{\kv}}(\sigma_{2}\sigma_{3}\sigma_{4}+2\sigma_{2})\biggl] \nnr
&\times \left(-\frac{1}{\beta}\right)\sum_{n}{\rm g}_{\kv,n,\sigma_{1}}{\rm g}_{\kv,n,\sigma_{2}}({\rm g}_{\kv,n+\ell,\sigma_{3}}{\rm g}_{\kv,n+\ell,\sigma_{4}}-{\rm g}_{\kv,n-\ell,\sigma_{3}}{\rm g}_{\kv,n-\ell,\sigma_{4}}), \nnr
g_3(i\Omega_{\ell}) &= -\frac{1}{2}\frac{J_{sd}}{2^4}\left(\frac{e}{m}\right)^2\sum_{\kv}\sum_{\sigma_{1}\sigma_{2}\sigma_{3}\sigma_{4}}
\sigma_{1}\sigma_{2}\left(\frac{\gamma_{\kv}}{\alpha_{\rm R}}\right)^2 \nnr
&\times \left(-\frac{1}{\beta}\right)\sum_{n}{\rm g}_{\kv,n,\sigma_{1}}{\rm g}_{\kv,n,\sigma_{2}}({\rm g}_{\kv,n+\ell,\sigma_{3}}{\rm g}_{\kv,n+\ell,\sigma_{4}}-{\rm g}_{\kv,n-\ell,\sigma_{3}}{\rm g}_{\kv,n-\ell,\sigma_{4}})
.
\end{align}

\section{Gauge transformation} \label{C}
Under the gauge transformation $A_{\mu}(\qv)\to A_{\mu}(\qv)+iq_{\mu}\Lambda(\qv)$, where $\Lambda(\qv)$ is the gauge degree of freedom, the change in the effective Hamiltonian linear in $\Lambda$ is given by
\begin{align}
\delta H_{\rm eff}&=-i\sum_{\qv}
\Lambda(\qv)
\begin{bmatrix}
g_{1}(\bm{{\cal A}}_{\rm R}\cdot\qv)[\qv\cdot\Av(-\qv)-\qv\cdot\Av(-\qv)] \\
+g_{2}
\begin{bmatrix}
(\bm{{\cal A}}_{\rm R}\cdot\Av(-\qv))\qv^2+(\qv\cdot\Av(-\qv))(\bm{{\cal A}}_{\rm R}\cdot\qv) \\
-(\bm{{\cal A}}_{\rm R}\cdot\qv)(\qv\cdot\Av(-\qv))-(\bm{{\cal A}}_{\rm R}\cdot\Av(-\qv))\qv^2
\end{bmatrix}
\\
+g_{3}[\Av(-\qv)\cdot(\alphaRv\times\qv)(\Mv^{\perp}\cdot\qv)-(\Mv^{\perp}\cdot\qv)\Av(-\qv)\cdot(\alphaRv\times\qv)]
\end{bmatrix}
.
\end{align}
\end{widetext}



\bibliographystyle{jplain}
\bibliography{161222temf}

\begin{thebibliography}{16}%
\makeatletter
\providecommand \@ifxundefined [1]{%
 \@ifx{#1\undefined}
}%
\providecommand \@ifnum [1]{%
 \ifnum #1\expandafter \@firstoftwo
 \else \expandafter \@secondoftwo
 \fi
}%
\providecommand \@ifx [1]{%
 \ifx #1\expandafter \@firstoftwo
 \else \expandafter \@secondoftwo
 \fi
}%
\providecommand \natexlab [1]{#1}%
\providecommand \enquote  [1]{``#1''}%
\providecommand \bibnamefont  [1]{#1}%
\providecommand \bibfnamefont [1]{#1}%
\providecommand \citenamefont [1]{#1}%
\providecommand \href@noop [0]{\@secondoftwo}%
\providecommand \href [0]{\begingroup \@sanitize@url \@href}%
\providecommand \@href[1]{\@@startlink{#1}\@@href}%
\providecommand \@@href[1]{\endgroup#1\@@endlink}%
\providecommand \@sanitize@url [0]{\catcode `\\12\catcode `\$12\catcode
  `\&12\catcode `\#12\catcode `\^12\catcode `\_12\catcode `\%12\relax}%
\providecommand \@@startlink[1]{}%
\providecommand \@@endlink[0]{}%
\providecommand \url  [0]{\begingroup\@sanitize@url \@url }%
\providecommand \@url [1]{\endgroup\@href {#1}{\urlprefix }}%
\providecommand \urlprefix  [0]{URL }%
\providecommand \Eprint [0]{\href }%
\providecommand \doibase [0]{http://dx.doi.org/}%
\providecommand \selectlanguage [0]{\@gobble}%
\providecommand \bibinfo  [0]{\@secondoftwo}%
\providecommand \bibfield  [0]{\@secondoftwo}%
\providecommand \translation [1]{[#1]}%
\providecommand \BibitemOpen [0]{}%
\providecommand \bibitemStop [0]{}%
\providecommand \bibitemNoStop [0]{.\EOS\space}%
\providecommand \EOS [0]{\spacefactor3000\relax}%
\providecommand \BibitemShut  [1]{\csname bibitem#1\endcsname}%
\let\auto@bib@innerbib\@empty
\bibitem [{\citenamefont {Slonczewski}(1996)}]{Slonczewski96}%
  \BibitemOpen
  \bibfield  {author} {\bibinfo {author} {\bibfnamefont {J. C.}~\bibnamefont
  {Slonczewski}},\ }\href@noop {http://www.sciencedirect.com/science/article/pii/0304885396000625} {\bibfield  {journal} {\bibinfo  {journal} {J. Magn. Magn. Mater.}\ }\textbf {\bibinfo {volume} {159}},\ \bibinfo {pages}
  {L1} (\bibinfo {year} {1996})}\BibitemShut {NoStop}%
\bibitem [{\citenamefont {Berger}(1996)}]{Berger96}%
  \BibitemOpen
  \bibfield  {author} {\bibinfo {author} {\bibfnamefont {L.}~\bibnamefont
  {Berger}},\ }\href@noop {http://link.aps.org/doi/10.1103/PhysRevB.54.9353} {\bibfield  {journal} {\bibinfo  {journal} {Phys. Rev. B}\ }\textbf {\bibinfo {volume} {54}},\ \bibinfo {pages}
  {9353} (\bibinfo {year} {1996})}\BibitemShut {NoStop}%
\bibitem [{\citenamefont {Ast}\ \emph {et~al.}(2007)\citenamefont {Ast},
  \citenamefont {Henk}, \citenamefont {Ernst}, \citenamefont {Moreschini},
  \citenamefont {Falub}, \citenamefont {Pacil\'{e}}, \citenamefont {Bruno},
  \citenamefont {Kern},\ and\ \citenamefont {Grioni}}]{Ast07}%
  \BibitemOpen
  \bibfield  {author} {\bibinfo {author} {\bibfnamefont {C.~R.}\ \bibnamefont
  {Ast}}, \bibinfo {author} {\bibfnamefont {J.}~\bibnamefont {Henk}}, \bibinfo
  {author} {\bibfnamefont {A.}~\bibnamefont {Ernst}}, \bibinfo {author}
  {\bibfnamefont {L.}~\bibnamefont {Moreschini}}, \bibinfo {author}
  {\bibfnamefont {M.~C.}\ \bibnamefont {Falub}}, \bibinfo {author}
  {\bibfnamefont {D.}~\bibnamefont {Pacil\'{e}}}, \bibinfo {author}
  {\bibfnamefont {P.}~\bibnamefont {Bruno}}, \bibinfo {author} {\bibfnamefont
  {K.}~\bibnamefont {Kern}}, \ and\ \bibinfo {author} {\bibfnamefont
  {M.}~\bibnamefont {Grioni}},\ }\href {\doibase 10.1103/PhysRevLett.98.186807}
  {\bibfield  {journal} {\bibinfo  {journal} {Phys. Rev. Lett.}\ }\textbf
  {\bibinfo {volume} {98}},\ \bibinfo {eid} {186807} (\bibinfo {year}
  {2007})}\BibitemShut {NoStop}%
\bibitem [{\citenamefont {Rashba}(1960)}]{Rashba60}%
  \BibitemOpen
  \bibfield  {author} {\bibinfo {author} {\bibfnamefont {E. I.}~\bibnamefont
  {Rashba}},\ }\href@noop {} {\bibfield  {journal} {\bibinfo  {journal} {Sov.
  Phys. Solid State}\ }\textbf {\bibinfo {volume} {2}},\ \bibinfo {pages}
  {1109} (\bibinfo {year} {1960})}\BibitemShut {NoStop}%
\bibitem [{\citenamefont {Edelstein}(1990)}]{Edelstein90}%
  \BibitemOpen
  \bibfield  {author} {\bibinfo {author} {\bibfnamefont {V.}~\bibnamefont
  {Edelstein}},\ }\href {\doibase 10.1016/0038-1098(90)90963-C} {\bibfield
  {journal} {\bibinfo  {journal} {Solid State Commun.}\ }\textbf
  {\bibinfo {volume} {73}},\ \bibinfo {pages} {233 } (\bibinfo {year}
  {1990})}\BibitemShut {NoStop}%
\bibitem [{\citenamefont {Shen}\ \emph {et~al.}(2014)\citenamefont {Shen},
  \citenamefont {Vignale}, \ and\ \citenamefont {Raimondi}}]{Shen14}%
  \BibitemOpen
  \bibfield  {author} {\bibinfo {author} {\bibfnamefont {K.}\ \bibnamefont
  {Shen}}, \bibinfo {author} {\bibfnamefont {G.}\ \bibnamefont {Vignale}},
  \ and\
  \bibinfo {author} {\bibfnamefont {R.}\ \bibnamefont {Raimondi}},\ }\href
  {http://link.aps.org/doi/10.1103/PhysRevLett.112.096601} {\bibfield  {journal} {\bibinfo
  {journal} {Phys. Rev. Lett.}\ }\textbf {\bibinfo {volume} {112}},\ \bibinfo
  {pages} {096601} (\bibinfo {year} {2014})}\BibitemShut {NoStop}%
\bibitem [{\citenamefont {Sanchez}\ \emph {et~al.}(2013)\citenamefont
  {Sanchez}, \citenamefont {Vila}, \citenamefont {Desfonds}, \citenamefont
  {Gambarelli}, \citenamefont {Attane}, \citenamefont {De~Teresa},
  \citenamefont {Magen},\ and\ \citenamefont {Fert}}]{Sanchez13}%
  \BibitemOpen
  \bibfield  {author} {\bibinfo {author} {\bibfnamefont {J.~C.~R.}\
  \bibnamefont {S{\'a}nchez}}, \bibinfo {author} {\bibfnamefont {L.}~\bibnamefont
  {Vila}}, \bibinfo {author} {\bibfnamefont {G.}~\bibnamefont {Desfonds}},
  \bibinfo {author} {\bibfnamefont {S.}~\bibnamefont {Gambarelli}}, \bibinfo
  {author} {\bibfnamefont {J.~P.}\ \bibnamefont {Attan{\'e}}}, \bibinfo {author}
  {\bibfnamefont {J.~M.}\ \bibnamefont {De~Teresa}}, \bibinfo {author}
  {\bibfnamefont {C.}~\bibnamefont {Mag{\'e}n}}, \ and\ \bibinfo {author}
  {\bibfnamefont {A.}~\bibnamefont {Fert}},\ }\href
  {http://dx.doi.org/10.1038/ncomms3944} {\bibfield  {journal} {\bibinfo
  {journal} {Nat Commun}\ }\textbf {\bibinfo {volume} {4}},\ \bibinfo {pages}
  {2944} (\bibinfo {year} {2013})}\BibitemShut
  {NoStop}%
\bibitem [{\citenamefont {Shibata}\ \emph {et~al.}(2016)\citenamefont
  {Shibata}, \citenamefont {Takeuchi}, \citenamefont {Kohno},\ and\
  \citenamefont {Tatara}}]{Shibata16}%
  \BibitemOpen
  \bibfield  {author} {\bibinfo {author} {\bibfnamefont {J.}~\bibnamefont
  {Shibata}}, \bibinfo {author} {\bibfnamefont {A.}~\bibnamefont {Takeuchi}},
  \bibinfo {author} {\bibfnamefont {H.}~\bibnamefont {Kohno}}, \ and\ \bibinfo
  {author} {\bibfnamefont {G.}~\bibnamefont {Tatara}},\ }\href {\doibase
  10.7566/JPSJ.85.033701} {\bibfield  {journal} {\bibinfo  {journal} {J. Phys. Soc. Jpn.}\ }\textbf {\bibinfo {volume} {85}},\
  \bibinfo {pages} {033701} (\bibinfo {year} {2016})} \BibitemShut {NoStop}%
\bibitem [{\citenamefont {Narimanov}\ and\ \citenamefont
  {Kildishev}(2015)}]{Narimanov15}%
  \BibitemOpen
  \bibfield  {author} {\bibinfo {author} {\bibfnamefont {E.~E.}\ \bibnamefont
  {Narimanov}}\ and\ \bibinfo {author} {\bibfnamefont {A.~V.}\ \bibnamefont
  {Kildishev}},\ }\href {http://dx.doi.org/10.1038/nphoton.2015.56} {\bibfield
  {journal} {\bibinfo  {journal} {Nat Photon}\ }\textbf {\bibinfo {volume}
  {9}},\ \bibinfo {pages} {214} (\bibinfo {year} {2015})}\BibitemShut {NoStop}%
\bibitem [{\citenamefont {Takeuchi}\ and\ \citenamefont
  {Tatara}(2012)}]{Takeuchi12}%
  \BibitemOpen
  \bibfield  {author} {\bibinfo {author} {\bibfnamefont {A.}~\bibnamefont
  {Takeuchi}}\ and\ \bibinfo {author} {\bibfnamefont {G.}~\bibnamefont
  {Tatara}},\ }\href {\doibase 10.1143/JPSJ.81.033705} {\bibfield  {journal}
  {\bibinfo  {journal} {J. Phys. Soc. Jpn.}\ }\textbf
  {\bibinfo {volume} {81}},\ \bibinfo {pages} {033705} (\bibinfo {year}
  {2012})}\BibitemShut {NoStop}%
\bibitem [{\citenamefont {Kim}\ \emph {et~al.}(2012)\citenamefont {Kim},
  \citenamefont {Moon}, \citenamefont {Lee},\ and\ \citenamefont
  {Lee}}]{Kim12}%
  \BibitemOpen
  \bibfield  {author} {\bibinfo {author} {\bibfnamefont {K.-W.}\ \bibnamefont
  {Kim}}, \bibinfo {author} {\bibfnamefont {J.-H.}\ \bibnamefont {Moon}},
  \bibinfo {author} {\bibfnamefont {K.-J.}\ \bibnamefont {Lee}}, \ and\
  \bibinfo {author} {\bibfnamefont {H.-W.}\ \bibnamefont {Lee}},\ }\href
  {\doibase 10.1103/PhysRevLett.108.217202} {\bibfield  {journal} {\bibinfo
  {journal} {Phys. Rev. Lett.}\ }\textbf {\bibinfo {volume} {108}},\ \bibinfo
  {pages} {217202} (\bibinfo {year} {2012})}\BibitemShut {NoStop}%
\bibitem [{\citenamefont {Nakabayashi}\ and\ \citenamefont
  {Tatara}(2014)}]{Nakabayashi14}%
  \BibitemOpen
  \bibfield  {author} {\bibinfo {author} {\bibfnamefont {N.}~\bibnamefont
  {Nakabayashi}}\ and\ \bibinfo {author} {\bibfnamefont {G.}~\bibnamefont
  {Tatara}},\ }\href {http://stacks.iop.org/1367-2630/16/i=1/a=015016}
  {\bibfield  {journal} {\bibinfo  {journal} {New J. Phys.}\ }\textbf
  {\bibinfo {volume} {16}},\ \bibinfo {pages} {015016} (\bibinfo {year}
  {2014})}\BibitemShut {NoStop}%
\bibitem [{\citenamefont {Hayami}\ \emph {et~al.}(2014)\citenamefont {Hayami},
  \citenamefont {Kusunose},\ and\ \citenamefont {Motome}}]{Hayami14}%
  \BibitemOpen
  \bibfield  {author} {\bibinfo {author} {\bibfnamefont {S.}~\bibnamefont
  {Hayami}}, \bibinfo {author} {\bibfnamefont {H.}~\bibnamefont {Kusunose}}, \
  and\ \bibinfo {author} {\bibfnamefont {Y.}~\bibnamefont {Motome}},\ }\href
  {\doibase 10.1103/PhysRevB.90.024432} {\bibfield  {journal} {\bibinfo
  {journal} {Phys. Rev. B}\ }\textbf {\bibinfo {volume} {90}},\ \bibinfo
  {pages} {024432} (\bibinfo {year} {2014})}\BibitemShut {NoStop}%
\bibitem [{\citenamefont {Sawada}\ and\ \citenamefont
  {Nagaosa}(2005)}]{Sawada05}%
  \BibitemOpen
  \bibfield  {author} {\bibinfo {author} {\bibfnamefont {K.}~\bibnamefont
  {Sawada}}\ and\ \bibinfo {author} {\bibfnamefont {N.}~\bibnamefont
  {Nagaosa}},\ }\href {\doibase 10.1103/PhysRevLett.95.237402} {\bibfield
  {journal} {\bibinfo  {journal} {Phys. Rev. Lett.}\ }\textbf {\bibinfo
  {volume} {95}},\ \bibinfo {pages} {237402} (\bibinfo {year}
  {2005})}\BibitemShut {NoStop}%
\bibitem [{\citenamefont {Proskurin}\ \emph {et~al.}(2016)\citenamefont
  {Proskurin}, \citenamefont {Ovchinnikov}, \citenamefont {Nosov},\ and\
  \citenamefont {Kishine}}]{Proskurin16}%
  \BibitemOpen
  \bibfield  {author} {\bibinfo {author} {\bibfnamefont {I.}~\bibnamefont
  {Proskurin}}, \bibinfo {author} {\bibfnamefont {A.~S.}~\bibnamefont {Ovchinnikov}},
  \bibinfo {author} {\bibfnamefont {P.}~\bibnamefont {Nosov}}, \ and\ \bibinfo
  {author} {\bibfnamefont {J.-i.}~\bibnamefont {Kishine}},\ }\href {\doibase
  } {\bibfield  {journal} {\bibinfo  {journal} {arXiv:1608.02722}\ } \textbf {\bibinfo {volume} {}}\
  \bibinfo {pages} {}(\bibinfo {year} {2016})} \BibitemShut {NoStop}%
\bibitem [{\citenamefont {Volovik}(1987)}]{Volovik87}%
  \BibitemOpen
  \bibfield  {author} {\bibinfo {author} {\bibfnamefont {G. E.}\ \bibnamefont
  {Volovik}},\ }\href@noop {} {\bibfield  {journal} {\bibinfo  {journal} {J.
  Phys. C: Solid State Phys.}\ }\textbf {\bibinfo {volume} {20}},\ \bibinfo
  {pages} {L83} (\bibinfo {year} {1987})}\BibitemShut {NoStop}%
\bibitem [{\citenamefont {Tatara}\ and\ \citenamefont {Nakabayashi}(2014)}]{GT14}%
  \BibitemOpen
  \bibfield  {author} {\bibinfo {author} {\bibfnamefont {G.}~\bibnamefont
  {Tatara}}\ and\ \bibinfo {author} {\bibfnamefont {N.}~\bibnamefont {Nakabayashi}},\
  }\href {http://dx.doi.org/10.1063/1.4870919} {\bibfield  {journal}
  {\bibinfo  {journal} {J. Appl. Phys.}\ }\textbf {\bibinfo {volume} {115}},\
  \bibinfo {eid} {172609} (\bibinfo {year} {2014})}\BibitemShut {NoStop}%
\bibitem [{\citenamefont {Kohno}\ \emph {et~al.}(2009)\citenamefont {Kohno},
  \citenamefont {Kawabata}, \citenamefont {Noguchi}, \citenamefont {Ueta},
  \citenamefont {Shibata},\ and\ \citenamefont {Tatara}}]{Kohno09}%
  \BibitemOpen
  \bibfield  {author} {\bibinfo {author} {\bibfnamefont {H.}~\bibnamefont
  {Kohno}}, \bibinfo {author} {\bibfnamefont {S.}~\bibnamefont {Kawabata}},
  \bibinfo {author} {\bibfnamefont {T.}~\bibnamefont {Noguchi}}, \bibinfo
  {author} {\bibfnamefont {S.}~\bibnamefont {Ueta}}, \bibinfo {author}
  {\bibfnamefont {J.}~\bibnamefont {Shibata}}, \ and\ \bibinfo {author}
  {\bibfnamefont {G.}~\bibnamefont {Tatara}},\ }\href@noop {} {\bibfield
  {journal} {\bibinfo  {journal} {in {\textit{Proceedings of ISQM-Tokyo '08}}}},
  \bibinfo {pages} {pp.111-117}},\ {Eds. S. Ishioka and K. Fujikawa},
  \ (\bibinfo  {publisher} {World Scientific, Singapore},\ \bibinfo {year} {2009  })\ \bibinfo {note} {(see also arXiv:0912.1676)}
  \BibitemShut {NoStop}%
\bibitem [{\citenamefont {Kawaguchi}\ and\ \citenamefont {Tatara}(2014)}]{Kawaguchi14}
  \bibfield  {author} {\bibinfo {author} {\bibfnamefont {H.}~\bibnamefont
  {Kawaguchi}}\ and\ \bibinfo {author} {\bibfnamefont {G.}~\bibnamefont {Tatara}  },\
  }\href {http://dx.doi.org/10.7566/JPSJ.83.074710} {\bibfield  {journal}
  {\bibinfo  {journal} {J. Phys. Soc. Jpn.}\ }\textbf {\bibinfo {volume} {83}},\  \bibinfo {eid} {074710} (\bibinfo {year} {2014})}\BibitemShut {NoStop}%
\bibitem [{\citenamefont {Taguchi}\ \emph {et~al.}(2012)\citenamefont
  {Taguchi}, \citenamefont {Ohe},\ and\ \citenamefont {Tatara}}]{Taguchi12}%
  \BibitemOpen
  \bibfield  {author} {\bibinfo {author} {\bibfnamefont {K.}~\bibnamefont
  {Taguchi}}, \bibinfo {author} {\bibfnamefont {J.-i.}\ \bibnamefont {Ohe}}, \
  and\ \bibinfo {author} {\bibfnamefont {G.}~\bibnamefont {Tatara}},\ }\href
  {\doibase 10.1103/PhysRevLett.109.127204} {\bibfield  {journal} {\bibinfo
  {journal} {Phys. Rev. Lett.}\ }\textbf {\bibinfo {volume} {109}},\ \bibinfo
  {pages} {127204} (\bibinfo {year} {2012})}\BibitemShut {NoStop}%
\bibitem [{\citenamefont {Qi}\ and\ \citenamefont {Zhang}(2011)}]{Qi11}%
  \BibitemOpen
  \bibfield  {author} {\bibinfo {author} {\bibfnamefont {X.-L.}\ \bibnamefont
  {Qi}}\ and\ \bibinfo {author} {\bibfnamefont {S.-C.}\ \bibnamefont {Zhang}},\
  }\href {\doibase 10.1103/RevModPhys.83.1057} {\bibfield  {journal} {\bibinfo
  {journal} {Rev. Mod. Phys.}\ }\textbf {\bibinfo {volume} {83}},\ \bibinfo
  {pages} {1057} (\bibinfo {year} {2011})}\BibitemShut {NoStop}%
\bibitem [{\citenamefont {Jackson}(1998)}]{Jackson98}%
  \BibitemOpen
  \bibfield  {author} {\bibinfo {author} {\bibfnamefont {J.~D.}\ \bibnamefont
  {Jackson}},\ }\href@noop {} {\emph {\bibinfo {title} {Classical
  Electrodynamics}}},\ \bibinfo {edition} {3rd}\ ed.\ (\bibinfo
  {publisher} {Wiley, New York},\ \bibinfo {year} {1998})\BibitemShut {NoStop}%
\bibitem [{\citenamefont {Leonhardt}\ and\ \citenamefont
  {Piwnicki}(1999)}]{Leonhardt99}%
  \BibitemOpen
  \bibfield  {author} {\bibinfo {author} {\bibfnamefont {U.}~\bibnamefont
  {Leonhardt}}\ and\ \bibinfo {author} {\bibfnamefont {P.}~\bibnamefont
  {Piwnicki}},\ }\href {http://link.aps.org/doi/10.1103/PhysRevA.60.4301} {\bibfield
  {journal} {\bibinfo  {journal} {Phys. Rev. A}\ }\textbf {\bibinfo
  {volume} {60}},\ \bibinfo {pages} {4301} (\bibinfo {year}
  {1999})}\BibitemShut {NoStop}%
\bibitem [{\citenamefont {Tatara}\ \emph {et~al.}(2008)\citenamefont {Tatara},
  \citenamefont {Kohno},\ and\ \citenamefont {Shibata}}]{TKS_PR08}%
  \BibitemOpen
  \bibfield  {author} {\bibinfo {author} {\bibfnamefont {G.}~\bibnamefont
  {Tatara}}, \bibinfo {author} {\bibfnamefont {H.}~\bibnamefont {Kohno}}, \
  and\ \bibinfo {author} {\bibfnamefont {J.}~\bibnamefont {Shibata}},\ }\href
  {\doibase doi:10.1016/j.physrep.2008.07.003} {\bibfield  {journal} {\bibinfo
  {journal} {Phys. Rep.}\ }\textbf {\bibinfo {volume} {468}},\ \bibinfo
  {pages} {213} (\bibinfo {year} {2008})}\BibitemShut {NoStop}%
\bibitem [{\citenamefont {Phuc}\ \emph {et~al.}(2015)\citenamefont
  {Phuc}, \citenamefont {Tatara}, \citenamefont {Kawaguchi},\ and\
  \citenamefont {Ueda}}]{Phuc15}%
  \BibitemOpen
  \bibfield  {author} {\bibinfo {author} {\bibfnamefont {N. T.}~\bibnamefont
  {Phuc}}, \bibinfo {author} {\bibfnamefont {G.}~\bibnamefont {Tatara}},
  \bibinfo {author} {\bibfnamefont {Y.}~\bibnamefont {Kawaguchi}}, \ and\ \bibinfo
  {author} {\bibfnamefont {M.}~\bibnamefont {Ueda}},\ }\href {http://dx.doi.org/10.1038/ncomms9135} {\bibfield  {journal} {\bibinfo  {journal}
  {Nat Commun}\ }\textbf {\bibinfo {volume} {6}},\ \bibinfo {pages}
  {8135} (\bibinfo {year} {2015})}\BibitemShut {NoStop}%
\bibitem [{\citenamefont {Kikuchi}\ \emph {et~al.}(2016)\citenamefont
  {Kikuchi}, \citenamefont {Koretsune}, \citenamefont {Arita},\ and\
  \citenamefont {Tatara}}]{Kikuchi16}%
  \BibitemOpen
  \bibfield  {author} {\bibinfo {author} {\bibfnamefont {T.}~\bibnamefont
  {Kikuchi}}, \bibinfo {author} {\bibfnamefont {T.}~\bibnamefont {Koretsune}},
  \bibinfo {author} {\bibfnamefont {R.}~\bibnamefont {Arita}}, \ and\ \bibinfo
  {author} {\bibfnamefont {G.}~\bibnamefont {Tatara}},\ }\href {\doibase
  10.1103/PhysRevLett.116.247201} {\bibfield  {journal} {\bibinfo  {journal}
  {Phys. Rev. Lett.}\ }\textbf {\bibinfo {volume} {116}},\ \bibinfo {pages}
  {247201} (\bibinfo {year} {2016})}\BibitemShut {NoStop}%
\bibitem [{\citenamefont {Spaldin}\ \emph {et~al.}(2008)\citenamefont {Spaldin}
  \citenamefont {Fiebig}, \ and\ \citenamefont{Mostovoy}}]{Spaldin08}%
  \BibitemOpen
  \bibfield  {author} {\bibinfo {author} {\bibfnamefont {N. A.}\ \bibnamefont
  {Spaldin}}, \bibinfo {author} {\bibfnamefont {M.}\ \bibnamefont {Fiebig}}, \ and\
  \bibinfo {author} {\bibfnamefont {M.}\ \bibnamefont {Mostovoy}},\ }\href
  {\doibase http://stacks.iop.org/0953-8984/20/i=43/a=434203} {\bibfield  {journal} {\bibinfo
  {journal} {J. Phys.: Condens. Matter}\ }\textbf {\bibinfo {volume} {20}},\ \bibinfo
  {pages} {434203} (\bibinfo {year} {2008})}\BibitemShut {NoStop}%
\bibitem [{\citenamefont {Baranova}\ \emph {et~al.}(1977)\citenamefont {Baranova},
  \citenamefont {Bogdanov},\ and\ \citenamefont {Zel'dovich}}]{Baranova77}%
  \BibitemOpen
  \bibfield  {author} {\bibinfo {author} {\bibfnamefont {N. B.}~\bibnamefont
  {Baranova}}, \bibinfo {author} {\bibfnamefont {Yu. V.}~\bibnamefont {Bogdanov}}, \
  and\ \bibinfo {author} {\bibfnamefont {B. Ya.}~\bibnamefont {Zei'dovich}},\ }\href
  {\doibase http://stacks.iop.org/0038-5670/20/i=10/a=R07} {\bibfield  {journal} {\bibinfo
  {journal} {Sov. Phys. Usp.}\ }\textbf {\bibinfo {volume} {20}},\ \bibinfo
  {pages} {870} (\bibinfo {year} {1977})}\BibitemShut {NoStop}%
\bibitem [{\citenamefont {Nakano}\ and\ \citenamefont
  {Kimura}(1969)}]{Nakano69}%
  \BibitemOpen
  \bibfield  {author} {\bibinfo {author} {\bibfnamefont {H.}~\bibnamefont
  {Nakano}}\ and\ \bibinfo {author} {\bibfnamefont {H.}~\bibnamefont
  {Kimura}},\ }\href {http://dx.doi.org/10.1143/JPSJ.27.519}
  {\bibfield  {journal} {\bibinfo  {journal} {J. Phys. Soc. Jpn.}\ }\textbf
  {\bibinfo {volume} {27}},\ \bibinfo {pages} {519} (\bibinfo {year}
  {1969})}\BibitemShut {NoStop}%
\bibitem [{\citenamefont {Mondal}\ \emph {et~al.}(2015)\citenamefont {Mondal},
  \citenamefont {Berritta}, \citenamefont {Paillard}, \citenamefont {Singh},
  \citenamefont {Dkhil}, \citenamefont {Oppeneer},
  ,\ and\ \citenamefont {Bellaiche}}]{Mondal15}%
  \BibitemOpen
  \bibfield  {author} {\bibinfo {author} {\bibfnamefont {R.}\ \bibnamefont
  {Mondal}}, \bibinfo {author} {\bibfnamefont {M.}~\bibnamefont {Berritta}}, \bibinfo
  {author} {\bibfnamefont {C.}~\bibnamefont {Paillard}}, \bibinfo {author}
  {\bibfnamefont {S.}~\bibnamefont {Singh}}, \bibinfo {author}
  {\bibfnamefont {B.}\ \bibnamefont {Dkhil}}, \bibinfo {author}
  {\bibfnamefont {P. M.}~\bibnamefont {Oppeneer}},
  \ and\ \bibinfo {author} {\bibfnamefont
  {L.}~\bibnamefont {Bellaiche}},\ }\href {http://link.aps.org/doi/10.1103/PhysRevB.92.100402}
  {\bibfield  {journal} {\bibinfo  {journal} {Phys. Rev. B}\ }\textbf
  {\bibinfo {volume} {92}},\ \bibinfo {eid} {100402} (\bibinfo {year}
  {2015})}\BibitemShut {NoStop}%
\bibitem [{\citenamefont {Train}\ \emph {et~al.}(2008)\citenamefont {Train},
  \citenamefont {Gheorghe}, \citenamefont {Krstic}, \citenamefont {Chamoreau},
  \citenamefont {Ovanesyan}, \citenamefont {Rikken}, \citenamefont {Gruselle},
  ,\ and\ \citenamefont {Verdaguer}}]{Train08}%
  \BibitemOpen
  \bibfield  {author} {\bibinfo {author} {\bibfnamefont {C.}\ \bibnamefont
  {Train}}, \bibinfo {author} {\bibfnamefont {R.}~\bibnamefont {Gheorghe}}, \bibinfo
  {author} {\bibfnamefont {V.}~\bibnamefont {Krstic}}, \bibinfo {author}
  {\bibfnamefont {L.-M.}~\bibnamefont {Chamoreau}}, \bibinfo {author}
  {\bibfnamefont {N.~S.}\ \bibnamefont {Ovanesyan}}, \bibinfo {author}
  {\bibfnamefont {G. L. J. A.}~\bibnamefont {Rikken}}, \bibinfo {author}
  {\bibfnamefont {M.}~\bibnamefont {Gruselle}},
  \ and\ \bibinfo {author} {\bibfnamefont
  {M.}~\bibnamefont {Verdaguer}},\ }\href {http://dx.doi.org/10.1038/nmat2256}
  {\bibfield  {journal} {\bibinfo  {journal} {Nat Mater}\ }\textbf
  {\bibinfo {volume} {7}},\ \bibinfo {eid} {729} (\bibinfo {year}
  {2008})}\BibitemShut {NoStop}%
\bibitem [{\citenamefont {Sakita}(1985)}]{Sakita85}%
  \BibitemOpen
  \bibfield  {author} {\bibinfo {author} {\bibfnamefont {B.}~\bibnamefont
  {Sakita}},\ }\href@noop {} {\emph {\bibinfo {title} {Quantum theory of
  many-variable systems and fields}}}\ (\bibinfo  {publisher} {World
  Scientific, Singapore},\ \bibinfo {year} {1985})\BibitemShut {NoStop}%
\bibitem [{\citenamefont {Abrikosov}(1975)}]{ADG75}%
  \BibitemOpen
 \bibfield  {author} {\bibinfo {author} {\bibfnamefont {A. A.}~\bibnamefont
  {Abrikosov}}, \bibinfo {author} {\bibfnamefont {L. P.}~\bibnamefont {Gorkov}}, 
  \ and\ \bibinfo {author} {\bibfnamefont {I. E.}\ \bibnamefont
  {Dzialoshinskii}},\ 
  }\href@noop {} {\emph {\bibinfo {title} {Methods of Quantum Field Theory in Statistical Physics}}}\ (\bibinfo  {publisher} {Dover, New York},\ \bibinfo {year} {1975})\ \bibinfo {pages} {}\BibitemShut {NoStop}%
\bibitem [{\citenamefont {Altland}(2006)}]{AS06}%
  \BibitemOpen
  \bibfield  {author} {\bibinfo {author} {\bibfnamefont {A.}~\bibnamefont
  {Altland}}\ and\ \bibinfo {author} {\bibfnamefont {B.}~\bibnamefont {Simons}},\
  }\href@noop {} {\emph {\bibinfo {title} {Condensed Matter Field Theory}}},\ \bibinfo {edition} {2nd}\ ed.\ (\bibinfo
  {publisher} {Cambridge University Press, New York},\ \bibinfo {year} {2006})\BibitemShut {NoStop}%
\bibitem [{\citenamefont {Burkov}\ and\ \citenamefont
  {Balents}(2011)}]{Burkov11}%
  \BibitemOpen
  \bibfield  {author} {\bibinfo {author} {\bibfnamefont {A. A.}~\bibnamefont
  {Burkov}}\ and\ \bibinfo {author} {\bibfnamefont {L.}~\bibnamefont
  {Balents}},\ }\href {http://link.aps.org/doi/10.1103/PhysRevLett.107.127205} {\bibfield
  {journal} {\bibinfo  {journal} {Phys. Rev. Lett.}\ }\textbf {\bibinfo
  {volume} {107}},\ \bibinfo {pages} {127205} (\bibinfo {year}
  {2011})}\BibitemShut {NoStop}%
\bibitem [{\citenamefont {Vazifeh}\ and\ \citenamefont
  {Franz}(2013)}]{Vazifeh13}%
  \BibitemOpen
  \bibfield  {author} {\bibinfo {author} {\bibfnamefont {M. M.}~\bibnamefont
  {Vazifeh}}\ and\ \bibinfo {author} {\bibfnamefont {M.}~\bibnamefont
  {Franz}},\ }\href {http://link.aps.org/doi/10.1103/PhysRevLett.111.027201} {\bibfield
  {journal} {\bibinfo  {journal} {Phys. Rev. Lett.}\ }\textbf {\bibinfo
  {volume} {111}},\ \bibinfo {pages} {027201} (\bibinfo {year}
  {2013})}\BibitemShut {NoStop}%
\bibitem [{\citenamefont {Zyuzin}\ and\ \citenamefont
  {Burkov}(2012)}]{Zyuzin12}%
  \BibitemOpen
  \bibfield  {author} {\bibinfo {author} {\bibfnamefont {A. A.}~\bibnamefont
  {Zyuzin}}\ and\ \bibinfo {author} {\bibfnamefont {A. A.}~\bibnamefont
  {Burkov}},\ }\href {http://link.aps.org/doi/10.1103/PhysRevB.86.115133} {\bibfield
  {journal} {\bibinfo  {journal} {Phys. Rev. B}\ }\textbf {\bibinfo
  {volume} {86}},\ \bibinfo {pages} {115133} (\bibinfo {year}
  {2012})}\BibitemShut {NoStop}%
\bibitem [{\citenamefont {Goswami}\ and\ \citenamefont
  {Tewari}(2013)}]{Goswami13}%
  \BibitemOpen
  \bibfield  {author} {\bibinfo {author} {\bibfnamefont {P.}~\bibnamefont
  {Goswami}}\ and\ \bibinfo {author} {\bibfnamefont {S.}~\bibnamefont
  {Tewari}},\ }\href {http://link.aps.org/doi/10.1103/PhysRevB.88.245107} {\bibfield
  {journal} {\bibinfo  {journal} {Phys. Rev. B}\ }\textbf {\bibinfo
  {volume} {88}},\ \bibinfo {pages} {245107} (\bibinfo {year}
  {2013})}\BibitemShut {NoStop}%
\bibitem [{\citenamefont {Fujikawa}(2004)}]{Fujikawa04}%
  \BibitemOpen
  \bibfield  {author} {\bibinfo {author} {\bibfnamefont {K.}~\bibnamefont
  {Fujikawa}}\ and\ \bibinfo {author} {\bibfnamefont {H.}~\bibnamefont {Suzuki}},\
  }\href@noop {} {\emph {\bibinfo {title} {Path Integrals and Quantum Anomalies}}}\ \bibinfo {edition} (\bibinfo
  {publisher} {Oxford University Press, Oxford},\ \bibinfo {year} {2004})\BibitemShut {NoStop}%
\bibitem [{\citenamefont {Kargarian}\ \emph {et~al.}(2015)\citenamefont {Kargarian},
  \citenamefont {Randeria}, \ and\ \citenamefont {Trivedi}}]{Kargarian15}%
  \BibitemOpen
  \bibfield  {author} {\bibinfo {author} {\bibfnamefont {M.}\ \bibnamefont
  {Kargarian}}, \bibinfo {author} {\bibfnamefont {M.}\ \bibnamefont {Randeria}},
  \ and\
  \bibinfo {author} {\bibfnamefont {N.}\ \bibnamefont {Trivedi}},\ }\href
  {http://dx.doi.org/10.1038/srep12683} {\bibfield  {journal} {\bibinfo
  {journal} {Sci. Rep.}\ }\textbf {\bibinfo {volume} {5}},\ \bibinfo
  {pages} {12683} (\bibinfo {year} {2015})}\BibitemShut {NoStop}%
\end{thebibliography}%

\end{document}